\documentclass[12pt,titlepage]{article}
\usepackage{epsfig}
\def\Lp{{\ell^2}}
\def\beq{\begin{equation}}
\def\eeq{\end{equation}}
\def\bea{\begin{eqnarray}}
\def\eea{\end{eqnarray}}
\begin{document}
%\print{DIAS-STP-10-10}
%\title{The cosmological constant and the black hole equation of state}
\title{The cosmological constant and black hole thermodynamic potentials}

\author{{Brian P. Dolan}\\
{\small Department of Mathematical Physics, National University
of Ireland,}\\
{\small  Maynooth, Ireland}\\
{\small and}\\
{\small Dublin Institute for Advanced Studies,
10 Burlington Rd., Dublin, Ireland}\\
{\small e-mail: bdolan@stp.dias.ie}}

\maketitle
\begin{abstract}
The thermodynamics of black holes in various dimensions
are described in the presence of a negative
cosmological constant which is treated as a thermodynamic variable,
interpreted as a pressure in the equation of state.   The black hole
mass is then identified with the enthalpy, rather than the internal energy,
and heat capacities are calculated at constant
pressure not at constant volume.  The Euclidean action
is associated with a bridge equation for the Gibbs free energy
and not the Helmholtz free energy. Quantum corrections to the enthalpy
and the equation of state of the BTZ black hole are studied.

\bigskip

\ 

\noindent
\hbox{Report No. DIAS-STP-10-10}\hfill \hbox{PACS nos: 04.60.-m;  04.70.Dy}
\end{abstract}

\section{Introduction}

 The thermodynamics of black-holes has been an active and fascinating
area of research ever since the early papers of Beckenstein and
Hawking's derivation of the temperature associated with the event horizon, 
\cite{Bekenstein,Hawking}.  
In most treatments of black-hole thermodynamics the cosmological constant, 
$\Lambda$, is treated as a fixed parameter (possibly zero)
but it has been considered as a dynamical variable in
\cite{HenneauxTeitelboim,Teitelboim}
and it has further been suggested that it is better to consider $\Lambda$
as a thermodynamic variable, \cite{CCK,WWXD,Sekiwa,Larranaga,KRT}.
Physically $\Lambda$ is interpreted as a thermodynamic pressure in \cite{KRT},
consistent with the observation in \cite{Sekiwa} that the conjugate
thermodynamic variable is proportional to a volume. 
This naturally leads to a slightly different interpretation of
the black-hole mass than is usual in thermodynamic treatments, the
black-hole mass is equated with enthalpy, $H$, in \cite{KRT}
rather than the internal energy $E$, as is more usual.

 A black-hole with a positive cosmological constant has both
a cosmological event horizon and a black-hole event horizon,
these have different Hawking temperatures associated with them
in general which necessarily complicates any thermodynamical
treatment. We shall therefore focus on the case of a negative
cosmological constant in this work though, many of the conclusions
are applicable to the positive $\Lambda$ case.  The negative
$\Lambda$ case is of course of interest for studies on AdS/CFT
correspondence and the considerations here are likely to be
relevant to current attempts to model condensed matter systems
in $2+1$ dimensions using the boundary of $3+1$ dimensional
anti-de Sitter space, see {\it e.g.} \cite{Hartnoll}.  In particular quantities
calculated at constant $\Lambda$ correspond physically to constant
pressure: specific heats, for example, are specific heats at
constant pressure and not specific heats at constant volume and it
is the former that are more relevant to solid state applications.  

 In this work the idea that $\Lambda$ is a thermodynamic pressure,
and the conjugate variable a thermodynamic volume, 
is elaborated on and the thermodynamical structure
developed further.  The relation between various thermodynamical
potentials and the black-hole equation of state in anti-de Sitter
space is explored in detail.  \S 2 deals with the thermodynamic
potentials in the 4-dimensional case; \S 3 discusses the black-hole
equation of state and \S 4 the partition function. 
In \S 5 the discussion is extended to arbitrary
dimensions while \S 6 deals specifically with the 3-dimensional
case of the BTZ black-hole, for which quantum corrections to
the thermodynamical potential are known, at least perturbatively.
Finally \S 6 summarizes the main points and conclusions. 

\section{Enthalpy}
\label{sec:Enthalpy}

Consider a black hole with mass M in the presence of a cosmological
constant $\Lambda$.  The cosmological constant generates a pressure 
\beq \label{LambdaP}
P=-\frac{\Lambda}{8\pi G_N}\eeq
and has an energy density $\epsilon$ associated with it with $\varepsilon + P=0$,
{\it i.e.} the enthalpy density associated with $\Lambda$ is zero.
If the black hole has a volume $V$ the total energy contained in $V$ 
is
\[ E=M+\varepsilon V = M -PV \qquad \Rightarrow \qquad M = E + PV,\]
hence $M$ is most naturally associated with the enthalpy $H$ of the black hole
\[ H = E + PV. \]
It is not obvious what the volume of a black hole should be.  The na{\rm \"i}ve
identification of $V$ with the volume of a sphere with the radius
of the event horizon $r_h$ is too simplistic
since the radial co-ordinate is time-like inside the horizon and
$\frac {4\pi r_h^3} {3} $ is not the volume of any space-like section 
of space-time inside the horizon.
It is suggested in \cite{KRT} that $V$ be identified with
the volume excluded by the black hole horizon from a spatial slice
exterior to the black hole, giving the na{\rm\"i}ve result
$V=\frac {4\pi r_h^3} {3} $ but
from a more physically acceptable perspective.
For the moment we shall leave $V$ unspecified and determine it below
from thermodynamic considerations.

The natural variables for enthalpy are entropy and pressure, so we 
should view $M$ as a function of $S$ and $P$,
\[ M=H(S,P).\]
The functional form of $H$ is determined by the geometry together
with the Hawking relation, that entropy is one quarter of the horizon area
\beq \label{EntropyArea}
S= \frac {\pi r_h^2}{\hbar G_N}= \frac{\pi r_h^2}{\Lp},\eeq
where $l=\sqrt{\hbar G_N}$ is the Planck length.\footnote{In the ensuing
analysis we shall make factors of $G_N$ and $\hbar$ explicit
in order to exhibit clearly which aspects of the physics are classical
and which are quantum.  Note that $\Lp S=\pi r_h^2$ is a classical
quantity, $S$ diverges in the classical limit while $\Lp S$
remains finite.}

The metric of four-dimensional space-time is given by
\[ d^2s = -f(r) dt^2 + f^{-1}(r)dr^2 + r^2 d\Omega^2, \]
with 
\beq \label{fdef}
f(r) = 1 -\frac{2G_N M}{r} - \frac{\Lambda}{3} r^2, 
\eeq
and $d\Omega^2=d\theta^2+\sin^2\theta d\phi^2$ the
solid angle area element.  The event horizon is defined by $f(r_h)=0$,
\beq \label{rh}
\frac \Lambda 3 r_h^3 - r_h + 2G_N M=0.
\eeq
One can solve the cubic equation to find $r_h(M,\Lambda)$ analytically,
but the explicit form will not be needed in the following. 
For $\Lambda>0$ and $0<3M\sqrt{\Lambda}G_N<1$ there are two event horizons
and the region of space-time outside the black hole horizon but
inside the de Sitter horizon lies between them.  They coincide
when $3M\sqrt{\Lambda}G_N=1$. Each event horizon has a different Hawking
temperature associated with it and the system is not in thermal equilibrium.
In order to ensure thermal equilibrium we shall assume $\Lambda<0$
in the following.

Equation (\ref{fdef}) gives the mass as 
\beq M= \frac {r_h}{2G_N} \left(1  - \frac{\Lambda}{3} r_h^2 \right).
\label{Mass}
\eeq
Identifying $M$ with the enthalpy and using (\ref{LambdaP})
and (\ref{EntropyArea}) gives
\beq \label{Enthalpy}
H(S,P)=\frac {1} {2G_N} \left(\frac {\Lp S}{\pi}\right)^{\frac 1 2} 
\left(1+\frac {8 G_N \Lp S P} {3} \right).\eeq

The usual thermodynamic relations can now be used to determine the temperature
and the volume,
\bea \label{TH}
T&=&\left(\frac{\partial H}{\partial S} \right)_P\quad\Rightarrow\quad
T=\frac {\hbar} {4\pi}\left(\frac{\pi}{\Lp S }\right)^{\frac 1 2}
(1+8 P G_N \Lp S)=\frac{\hbar(1-\Lambda r_h^2)}
{4\pi r_h} \\
\label{VH}
V&=& \left(\frac{\partial H}{\partial P} \right)_S\quad
\Rightarrow\quad
V= \frac 4 3 \frac{(\Lp S)^{\frac 3 2}}{\sqrt{\pi}} = \frac {4 \pi r_h^3}{3}.
\eea
The Hawking temperature (\ref{TH}) comes as
no surprise since it follows from the usual formula
relating the temperature to the surface gravity $\kappa$ at the event horizon
\[ T=\frac {\hbar\kappa}{2\pi}=
 \frac{\hbar f'(r_h) }{4\pi}.  \]
Varying (\ref{rh}) gives
\[ df =  -\frac{2G_N}{r_h} d M   -\frac{r_h^2}{3} d\Lambda +f'(r_h)dr_h=0,\]
so
\[ \left.\frac{\partial M}{\partial r_h}\right|_\Lambda= \frac{r_h}{2 G}f'(r_h)
\qquad\Rightarrow\qquad
\left.\frac{\partial M}{\partial S}\right|_P= \frac{\hbar}{4\pi}f'(r_h)
,\]
which is equation (\ref{TH}).
Equation (\ref{VH}) suggests that the \lq\lq na{\rm\" i}ve'' volume is
indeed the correct one to use in thermodynamic relations.

Legendre transforming (\ref{Enthalpy}) gives the internal energy
\beq \label{Energy}
E(S,V)= H(S,P)-PV=\frac {1} {2 G_N} \sqrt{\frac {\Lp S}{\pi} },\eeq
but the Legendre transform is not invertible: because $H(S,P)$ is
linear in $P$, $E(S,V)$ is independent of $V$ and so the
pressure cannot be determined from a knowledge of $E(S,V)$ alone.
For the same reason $T=\left.\frac{\partial E}{\partial S}\right|_V$
gives the wrong answer for the temperature if the
pressure is non-zero.  

Nevertheless we can still use (\ref{Enthalpy}) to determine the heat capacity
at constant pressure using the standard thermodynamic relation
\[ 
C_P = \frac T {\left.\frac {\partial T}{\partial S}\right|_P}.
\]
One finds
\beq
C_P= 2S\left(\frac{8 G_N P \Lp S+1}{8 G_N P \Lp S-1}\right).\eeq
The heat capacity at constant volume
\[
C_V= \frac T {\left.\frac {\partial T}{\partial S}\right|_V}
= T \left.\frac {\partial S}{\partial T}\right|_V\]
vanishes, if we use the Hawking formula (\ref{EntropyArea}),
since the variation
of the entropy is necessarily zero when the volume, and hence $r_h$, is fixed.

It could have been anticipated in advance that $C_V=0$, from (\ref{Energy}):
$E$ depends only on $S$ and $S=
\frac{\pi}{\Lp} \left(\frac{3 V}{4\pi} \right)^{\frac 2 3}$ 
is a function of $V$ alone so $E$ is constant if $V$ is held fixed
as the temperature is varied.  

Local stability requires that $C_P>0$
so $8G_N P\Lp S>1$, or equivalently
\[ -\Lambda r_h^2>1, \]
so $\Lambda$ must be negative.  The fact that black holes can be 
thermodynamically stable in anti-de Sitter space-time is well known \cite{HP}.
Physically this condition for stability
can be understood as follows.  For $\Lambda<0$ the vacuum energy density
$\varepsilon<0$ so the black hole contains negative vacuum energy.  As
it radiates at constant pressure, and hence constant $\varepsilon$,
the volume decreases and the vacuum energy it contains increases
(becoming less negative). At the same time its temperature increases
hence the energy can go up as the temperature goes up,
if $|\Lambda|$ is large enough the heat capacity is positive and the black hole is 
stabilised by the negative vacuum energy.

 Local stability therefore implies a minimum temperature 
\beq \label{THP}
T_{min}=\hbar \sqrt{\frac{2G_N P}{\pi}}
\eeq
below which the black hole is not stable,
corresponding to the divergence in $C_P$ when 
$\left.\frac {\partial T}{\partial S}\right|_P=0$.
As is well known, this is below the Hawking-Page temperature,
$T_{HP}=\hbar\sqrt{\frac {8 G_N P} {3\pi}}$,
below which pure AdS space, with no black hole and $M=0$, has a lower free
energy than that of a black hole with the same $\Lambda$ and $M>0$
which occurs for $r_h=\sqrt{\frac 3 {|\Lambda|}}$, \cite{HP}.
 
\section{The black hole equation of state}

Writing equation (\ref{TH}) in terms of $V$ and $P$ gives the
black hole equation of state
\beq \label{EoS}
T(V,P)=\frac {\hbar}{4\pi}\left\{
\left(\frac {3V}{4\pi} \right)^{-\frac 1 3} + 
8\pi G_N P \left(\frac {3V}{4\pi} \right)^{\frac 1 3}
\right\}.
\eeq
For a given pressure there is a minimum volume at $T_{min}$,
\beq
V(T_{min})=\frac{4\pi}{3} \left( \frac 1 {8\pi G_N P}\right)^{\frac 3 2}.
\eeq

Figure 1 shows $T(V)$ for various pressures and figure 2 shows the
black hole indicator diagram, $P(V)$, for various temperatures.
The temperature as a function of entropy, at constant pressure, is
shown in figure 3, for comparison with the $J=Q=0$ case in figure 1 of
\cite{CCK}.  

\section{The partition function}

A key concept in understanding black hole thermodynamics is the
relation between the Euclidean path integral and the black hole
partition function \cite{GibbonsHawking}.
Defining the Euclidean action requires a regularisation procedure
as the volume of space-time is infinite and the Ricci scalar
is non-zero \cite{HP}.
A regularised Euclidean action can be obtained by adding surface terms
at large $r$ to cancel the infinities arising from taking $r\rightarrow\infty$
in the bulk integrals, \cite{HS,BK}.  Two terms are necessary,
one corresponding the extrinsic curvature of the sphere at large radius,
involving the unit normal $n^\mu$, and one simply proportional to the
area of the sphere, 
\begin{eqnarray}
I&=&-{1\over 16\pi G_N} \int_{\cal M} (R-2\Lambda)\sqrt{-g}\,d^4 x
\nonumber \\
&& +{1\over 8\pi G_N}\int_{\partial{\cal M}} \gamma^{\mu\nu}\nabla_\mu n_\nu \sqrt{-\gamma}\,d^3 x - {1\over 2\pi G_N L}\int_{\partial{\cal M}} \sqrt{-\gamma}\,d^3 x.
\label{IEintegral}
\end{eqnarray}
In \cite{GibbonsHawking} the integral is taken over $r_h<r<\infty$ with $\gamma$ the
three-dimensional metric on the asymptotic boundary $\partial{\cal M}$,
$r\rightarrow\infty$. In particular the event horizon is not considered to 
be part of $\partial{\cal M}$.
$L$ is the AdS length scale, $\Lambda =-\frac{3}{L^2}$, and
the Euclidean time parameter $x^0$ is periodic
with $0<x^0<\frac {1}{T}$.
Performing the integrals gives
\beq \label{IE}
I=\frac{r_h}{4G_N T}\left( 1 + \frac{\Lambda r_h^2}{3}\right). \eeq
In the Euclidean approach to quantum gravity, \cite{HP}, this is related to the partition function
$Z=e^{-I}$ through the bridge equation
\beq \label{bridge} TI=-T\ln Z,\eeq
and hence $TI$ is identified with the free energy 
\beq \label{Helmholtz} {\cal F}=\frac{r_h}{4G_N}\left( 1 + \frac{\Lambda r_h^2}{3}\right).\eeq
But is this the Helmholtz free energy $F(T,V)$ or the Gibbs free energy
$G(T,P)$?

The functional integral is performed with fixed $T$ and $\Lambda$,
so ${\cal F}(T,\Lambda)$ should be though of as a function of $T$ and $\Lambda$
and, using (\ref{TH}), it is readily shown that
\beq  d{\cal F}=-{\pi r_h^2 \over G_N \hbar} dT - {r_h^3 \over 6 G_N}d\Lambda.
\eeq
Hence
\beq -\left.{\partial {\cal F}\over \partial T}\right|_\Lambda= 
{\pi r_h^2 \over G_N \hbar} =S \qquad\hbox{and} \qquad 
\left.{\partial {\cal F}\over \partial P}\right|_T=
-8\pi G_N \left.{\partial {\cal F}\over \partial \Lambda}\right|_T = {4\pi r_h^3 \over 3}=V.
\eeq
These are the thermodynamic relations associated with  
the Gibbs free energy, $G(T,P)$, and not the Helmholtz free energy.
It is natural therefore to identify ${\cal F}=G(T,P)$ with
the Gibbs free energy.
The enthalpy is the Legendre transform of the Gibbs free energy
$H=G+TS$ and a simple calculation shows that $H=M$ is the
black-hole mass.

Euler's equation for thermodynamic potentials follows
from dimensional analysis.
Equation (\ref{rh}) is invariant under the rescalings
$r_h\rightarrow \eta r_h$, $\Lambda\rightarrow \eta^{-2}\Lambda$ 
and $M\rightarrow \eta M$ (keeping $G_N$ fixed).
Hence
\[ \eta M(S,P)=M(\eta^2 S,\eta^{-2} P)\qquad \Rightarrow 
\qquad M= 2(TS-PV),\]
which is easily checked.
This is Smarr's formula \cite{Smarr} treated from the same point of
view as in \cite{KRT}.  A simple consequence of this
scaling argument is that $M(S,P)$ must have the functional form
\[ M=\sqrt{S} \Phi(SP) \] for some function $\Phi(SP)$: 
in fact (\ref{Enthalpy}) shows that $\Phi(SP)$ is a linear function.
Corrections to the entropy can modify this simple scaling analysis.

\section{Higher dimensional black holes and different event horizon geometries}

In $D$ space-time dimensions the AdS-Schwarzschild line element is
\beq \label{line-element}
d^2s = - f(r)dt^2 + f^{-1}(r)dr^2 + r^d d\Omega_{(d)}^2,
\eeq
where $d=D-2$ and $d\Omega_{(d)}^2$ is the line element on a $d$-dimensional
sphere of unit radius.  Denoting the volume of the unit sphere 
by 
\[ \Omega_{(d)}= \frac{2\pi^{\frac d 2}}{\Gamma\left( \frac d 2\right)}\]
the $D$-dimensional Einstein equations give the function $f(r)$ to be
\beq \label{AdSf}
f(r)= 1 -\frac{16\pi G_N}{\Omega_{(d)}\, d} \frac{M}{r^{d-1}} -
\frac {\Lambda}{(d+1)} r^2,\eeq
with $\Lambda<0$ for AdS, \cite{Tangherlini}.

A more general line element is possible if surfaces of constant
$r$ are taken to be any Einstein space with constant curvature, such
as flat space or $d$-dimensional hyperbolic spaces.  Replacing the spherical
line element $d\Omega_{(d)}^2$ in (\ref{line-element}) by the appropriate
constant curvature line element in each case we
denote the line element by $d\Omega^2_{(d,k)}$ and the volume by 
$\Omega_{(d,k)}$.
For example $k=+1$ for spheres, $-1$ for hyperbolic space and $0$ for flat
space.  
Einstein's equations are now solved by replacing the function $f(r)$ 
with\footnote{For flat  space 
the volume can be taken to be finite by making
periodic identifications to get the topology of a torus.
For negatively curved spaces the same procedure gives more
complicated topologies, {\it e.g.} higher genus surfaces for $d=2$.}
\beq f(r)= k -\frac{16\pi G_N}{\Omega_{(d,k)}\, d} \frac{M}{r^{d-1}} -
\frac {\Lambda}{(d+1)} r^2.\eeq

For a general constant curvature Einstein space
with $d$-dimensional Ricci tensor
\[ R_{ij}=\lambda g_{ij}, \]
where $\lambda$ is a constant and $i,j=1,\ldots,d$,
the $d$-dimensional Ricci scalar is $R_{(d)}=\lambda d$ and
(\ref{line-element}) solves Einstein's equations with cosmological
constant $\Lambda$ provided 
\[ k=\frac{\lambda}{d-1}=\frac {R_{(d)}}{d(d-1)},\]
and $d\Omega^2_{(d,k)}$ corresponds to the line element of the
Einstein metric of the event horizon.
Note that $\Omega_{(d,k)}$ is dimensionless in the conventions adopted here,
the event horizon has area $r_h^d \Omega_{(d,k)}$ and $R_{(d)}$ is a 
dimensionless constant
fixed by the geometry of the event horizon, in particular it does not
depend on $r_h$.

The analysis of \S \ref{sec:Enthalpy} is easily repeated
with minor modifications.  In $D$ dimensions the Planck length
is given by $l^d= \hbar G_N$ and the entropy is
\beq \label{Entropyd}S=\frac {\Omega_{(d,k)}}{4} \frac{r_h^d}{l^d}.\eeq
Equating $M$ with the enthalpy and $\Lambda=-\frac{16\pi G_N}{d}P$ results 
in\footnote{We use the convention $R_{\mu\nu}=\Lambda g_{\mu\nu}$,
giving Einstein tensor 
$R_{\mu\nu}-{1\over 2} R \,g_{\mu\nu} = -\left({D-2\over 2}\right)\Lambda\, g_{\mu\nu}$.  Identifying $-\left({D-2\over 2}\right)\Lambda\, g_{\mu\nu}=8\pi G_N T_{\mu\nu}$, with $T_{\mu\nu}$ an energy-momentum tensor, leads to the 
quoted identification
$\Lambda=-{16 \pi G_N \over D-2}P$.  The gravitational action is 
$S_{Grav}=\frac 1 {16\pi G_N}\int\bigl(R-(D-2)\Lambda\bigr)\sqrt{-g}\,d^Dx$. }
\beq \label{Enthalpyd}
H(S,P)=\frac{\hbar S}{4\pi}
\left\{ 
\frac{R_{(d)}}{d-1}\left(\frac{4l^d S}{\Omega_{(d,k)}} \right)^{- \frac 1 d} +
\frac{16\pi G_NP}{d+1}
\left(\frac{4l^d S}{\Omega_{(d,k)}} \right)^{\frac 1 d}
\right\},
\eeq
from which all thermodynamic quantities can be calculated.

The thermodynamic volume is the na{\rm\" i}ve result
\beq \label{Volumed} V= \frac{\Omega_{(d,k)} r_h^{d+1}}{d+1},\eeq
and the equation of state is
\beq
T=\frac{\hbar}{4\pi d}\left\{ 
R_{(d)}\left(\frac{(d+1)V}{\Omega_{(d,k)}}\right)^{-\frac 1 {d+1}} + 
16\pi G_N P\left(\frac{(d+1)V}{\Omega_{(d,k)}}\right)^{\frac 1 {d+1}}\right\}.
\eeq

For positive
$k$ there is a minimum temperature and volume for any given fixed pressure
\[T_{min}=\frac{2\hbar}{d}\sqrt{\frac{R_{(d)}G_N P}{\pi}},\qquad 
V(T_{min})=\frac{\Omega_{(d,k)}}{d+1}\left( \frac{R_{(d)}}{16\pi G_N P}\right)^{\frac{d+1}{2}} \]
and the heat capacity is
\[ C_P= S d \left\{\frac
{16\pi G_N P \left(\frac{4l^d S}{\Omega_{(d,k)}}\right)^{\frac 2 d}+R_{(d)}}
{16\pi G_N P \left(\frac{4l^d S}{\Omega_{(d,k)}}\right)^{\frac 2 d}-R_{(d)}}\right\},
\]
which diverges at $T_{min}$ and is negative for $T<T_{min}$.
There is thus a Hawking-Page phase transition for any positive
curvature event horizon in $D\ge 4$ dimensions, for spatially flat
event horizons the specific heat $C_P=Sd$ is always positive while
for $R_{(d)}<0$ there is no minimum temperature, but there is still
a minimum value of $|\Lambda|$ below which $C_P$ is negative.
In all cases one must have 
\[|\Lambda| > \frac{\bigl|R_{(d)}\bigr|}{d\, r_h^2}  \]
for a black hole to be stable in anti-de Sitter space-time.

\section{The BTZ black hole}

It is worthwhile studying the special case of the $2+1$-dimensional
BTZ black hole, not only because it is conceptually and mathematically
simpler than its higher dimensional cousins 
but also because higher order corrections
to the entropy are easier to calculate and not as uncertain.
For a review of BTZ black holes see \cite{Carlip}.
 
The BTZ black hole has line element
\[ d^2s=-f(r)dt^2 + f^{-1}(r)dr^2 + r^2d\phi^2.\]

For a non-rotating BTZ black hole
\[ f(r)=-8G_N M + \frac {r^2}{L^2}, \]
with cosmological constant $\Lambda=-\frac 2 {L^2}$
giving a pressure $P=\frac 1 {8\pi G_N L^2}$.

The black hole radius
\beq \label{rBTZ}
r_h = \sqrt{8G_N M}L \eeq
is immediate.
The event horizon is a circle and the entropy is one-quarter of the
circumference in Planck units,
\[ S=\frac{\pi r_h}{2 l},\]
where $l=\hbar G_N$ in three dimensions.
Identifying the mass with the enthalpy
then gives
\beq \label{BTZEnthalpy}
H(S,P)= \frac {4\Lp}{\pi} S^2 P, \eeq
from which
\bea \label{BTZT} 
T&=& \frac {8\Lp}{\pi} S P = \frac{\hbar r_h}{2\pi L^2}\\
\label {BTZV} 
V&=& \frac {4\Lp}{\pi} S^2 = \pi r_h^2,\eea
the standard results (for uniformity of notation 
the symbol $V$ is used for the thermodynamic \lq\lq volume'',
even though it is an area in three space-time dimensions).
The Gibbs free energy is 
\[ G=H-TS=\frac{4l^2 S^2 P}{\pi}-2M=-M.\]
$C_P$ is easily calculated, since $T$ is linear in $S$
we have ${\left.\frac {\partial T}{\partial S}\right|_P}=\frac T S$ so
\[ C_P = \frac T {\left.\frac {\partial T}{\partial S}\right|_P}=S>0. \]
There is no local instability, there is however a global instability
since ordinary 3-dimensional AdS, with $f(r)=1+\frac{r^2}{L^2}$
corresponding to $M=-\frac{1}{8G_N}$, has $T=0$ giving $G=H=-\frac{1}{8G_N}$,
and so has lower Gibbs
free energy than the $M=0$ black-hole, making it more stable.
This suggests a phase transition from a black-hole
AdS state to pure $AdS_3$ when the Gibbs free energies are equal, 
which happens for $M=\frac 1 {8G_N}$ at a temperature $T=\hbar\sqrt{\frac{2G_N P}{\pi}}$,  
\cite{HP3,Myung}.  

Using (\ref{BTZV}) to express $S$ in terms of $V$ we derive the BTZ equation of state
\[ P V^{\frac 1 2}=\frac{\sqrt{\pi}}{4 \Lp} T.\]
From (\ref{BTZEnthalpy}) one finds $H=PV$ so Legendre transforming gives $E=0$,
the BTZ internal energy vanishes classically.

The partition function for the BTZ black-hole, 
including quantum corrections to all orders in perturbation theory, 
is given in \cite{MW} (earlier attempts at calculating
corrections to the BTZ 
black-hole entropy can be found in \cite{Carlip,CTeit,GM,MS}).
To understand the structure fully it is necessary to start with the
rotating black-hole, with metric \cite{Carlip}
\beq ds^2= -f(r) dt^2  + 
\frac {1}{f(r)}dr^2+ r^2 \biggl(d\phi - \frac {4G_N J}{r^2} dt\biggr)^2
\;,\label{RotatingBTZ}\eeq
where 
\[ f(r)= \biggl( -8G_N M + \frac {r^2}{L^2} + \frac{16 G_N^2J^2}{r^2}\biggr)\]
and $J$ is the angular momentum, bounded above by $J\le ML$.
There are now two event horizons, and inner and an outer horizon
at $r_+$ and $r_-$ respectively, with
\beq r_\pm^2 =4G_N M L^2 \left\{ 1\pm \left[ 1 - \biggl(\frac
J{M L}\biggr)^2\right]^{\frac 12} \right\}.\label{rpm}\eeq
The Hawking temperature associated with (\ref{RotatingBTZ}) is
\beq T= \frac {f'(r_+)}{4\pi}=\frac{(r_+^2 - r_-^2)\hbar}{2\pi L^2 r_+}.\eeq

The entropy corrections described in \cite{MW} are best discussed
in the Euclidean formalism. Wick rotating the time parameter
sends $t\rightarrow -it_E$ and the angular momentum 
also rotates, $J\rightarrow iJ_E$. The right hand side of
(\ref{rpm}) then translates to
\beq 4G_N M L^2 \left\{ 1\pm \left[ 1 + \biggl(
\frac {J_E}{M L}\biggr)^2\right]^{\frac 12} \right\}.\eeq
in the Euclidean sector and we see that $r_+\rightarrow r_{E,+}$
while $r_-\rightarrow ir_{E,-}$ where
\beq r_{E,\pm}^2 =4G_N M L^2\left\{ \left[1+\biggl(\frac
{J_E}{M L}\biggr)^2\right]^{\frac 12} \pm 1 \right\}.\label{rpmE}\eeq
The partition function is elegantly described in terms of the
dimensionless complex parameter 
\beq \tau =\frac{r_{E,-} + i r_{E,+}}{L}\eeq
which has $Im(\tau)>0$. The inverse Hawking temperature in the Euclidean formalism is
given by
\beq \frac 1 {2\pi T}  =\frac{r_{E,+}}{(r_{E,+}^2 + r_{E,-}^2)}\frac{L^2}{\hbar}
= \frac L \hbar \left\{Im\left(-\frac 1 \tau\right)\right\}^{-1}.\eeq
The BTZ partition function given in \cite{MW}, including all perturbative
quantum corrections, is most succinctly written by defining 
$q=e^{2\pi i\tau}$ in terms of which
\beq Z_{BTZ}= (q\bar q)^{-\frac L {16 \hbar G_N}}\prod_{n=2}^\infty |1-q^n|^{-2}.
\label{PertZBTZ}\eeq
Equation (\ref{PertZBTZ}) does not include non-perturbative 
quantum corrections,
but it suffices to illustrate this discussion of corrections to 
thermodynamic quantities. 
(Note that $\tau$ in our notation is $-\frac 1 \tau$ in the notation
of \cite{MW}, and $Z_{BTZ}$ here is denoted $Z_{1,0}$ there).

We now specialise to the case of zero angular momentum, when
\[ T=\frac{r_{E,+}}{2\pi}\frac{\hbar}{L^2}=\frac{r_+}{2\pi}\frac{\hbar}{L^2},
\qquad \tau=\frac {2\pi i TL}{\hbar} = i\frac {r_+} L \qquad\hbox{and}
\qquad q=e^{-4\pi^2 \frac{LT}{\hbar}}.\]
The partition function in this case is
\beq
Z_{BTZ}= e^{\frac{\pi^2 T L^2}{2\hbar^2 G_N}}
\prod_{n=2}^\infty \Bigl(1-e^{-4\pi^2 n\frac{TL}{\hbar}}\Bigr)^{-2}.
\label{ZBTZ}
\eeq
Thermodynamic functions can immediately be read off.
Defining $x=\frac{TL}{\hbar}=\frac{r_+}{2\pi L}$ the Gibbs free energy is
\beq \label{BTZGibbsFreeEnergy}
G(T,P)=-T\ln Z_{BTZ}=-\frac{\pi^2 x^2}{2G_N}+2 T \sum_{n=2}^\infty \ln 
\Bigl(1-e^{-4\pi^2 n x} \Bigr),\eeq
where the first term on the right hand side is the classical result.

The entropy was calculated in \cite{MW} and the 
enthalpy can be determined using the standard formula $H(S,P)=G+TS$, giving
\beq H=\frac{\pi^2 x^2}{2G_N} - 
8\pi^2 x T \sum_{n=2}^\infty  \frac n {e^{4\pi^2 n x} -1},\eeq
but this is not expressed explicitly in terms of the natural variables
$S$ and $P=\frac{1}{8\pi G_N L^2}$, the $S$ dependence is only implicit.

The quantum corrections embodied in the logarithmic terms of 
(\ref{BTZGibbsFreeEnergy}) modify the thermodynamic volume,
\beq 
V=\left.\frac{\partial G}{\partial P}\right|_T=\pi r_h^2\left[
1-8\pi \left(\frac{G_N\hbar}{L}\right)\sum_{n=2}^\infty 
\frac n {e^{4\pi^2 nx}-1} \right].
\eeq
We see that the quantum corrections serve to reduce the volume below
its classical value.  Figure 4 plots the PV-diagram, quantum 
effects reduce the volume to zero at finite pressure, the effect being
more pronounced at lower temperatures.

\section{Conclusions}

Some consequences of the suggestion in \cite{KRT},
that the correct thermodynamic interpretation
of black-hole mass in the presence of a negative cosmological constant
is that it should be associated with enthalpy rather than the
more usual interpretation of internal energy, have been explored and
expanded upon.  The cosmological constant is treated as a thermodynamic
variable proportional to the pressure and the black-hole mass
is identified with the enthalpy rather than the internal energy.
The interpretation of many thermodynamic quantities
is modified in this approach: specific heats, for example,
are naturally calculated as
specific heats at constant pressure rather than at constant volume
and the Euclidean action gives the Gibbs free energy
and not the Helmholtz free energy.

Black-hole solutions of Einstein's equations in any dimension,
and with any Einstein manifold with constant scalar curvature
as event horizon, can easily be constructed
and the classical equation of state determined.
The Hawking-Page transition is manifest as a change in sign the
specific heat and is present for black-holes with an event horizon
with positive curvature in any space-time dimension greater than three.

Quantum corrections to the thermodynamics relations for the BTZ
black-hole in three dimensions have been derived from the
partition function in \cite{MW}, which includes corrections to
all orders in perturbation theory but does not include
non-perturbative corrections.  These corrections reduce
the volume at a given pressure and temperature, with a finite pressure
giving zero volume.

The considerations presented here
will have important implications for AdS/CFT approaches to condensed
matter systems, in which the specific heat at constant
pressure has greater significance than the specific heat
constant volume.
 
It is a pleasure to thank the Perimeter Institute, Waterloo, Canada
for hosting a visit during which this work was initiated.
This work was partly funded by the 
EU Research Training Network in Noncommutative Geometry (EU-NCG).

\pagebreak

\centerline{\includegraphics{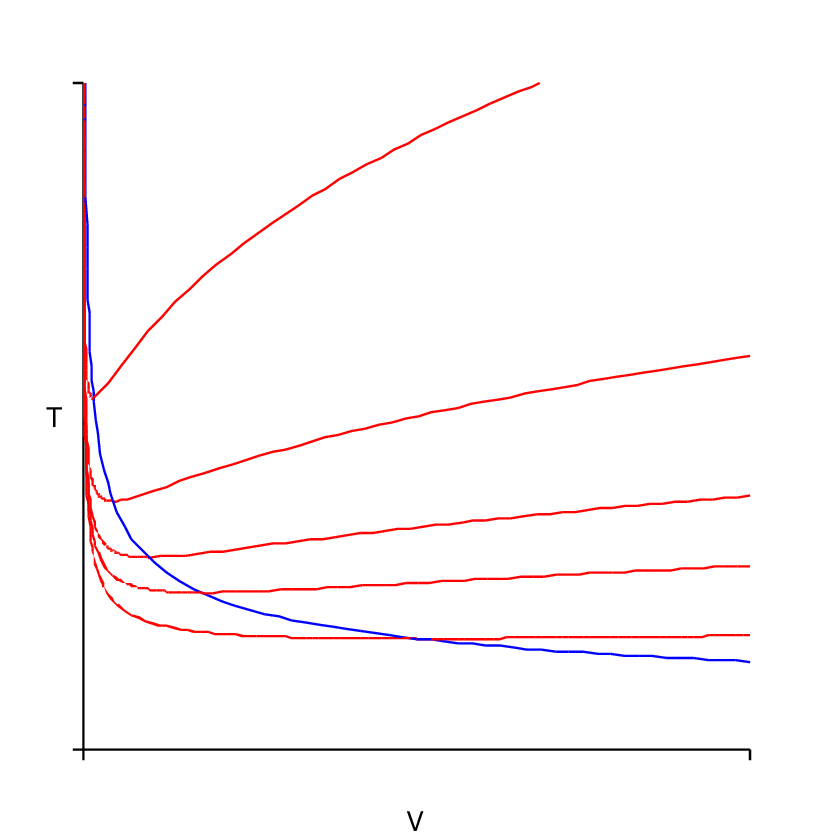}}

\vskip 10cm

\centerline{Figure 1: Black hole T-V diagram, showing curves of 
constant pressure}
\centerline{in AdS.  The blue line shows the stability limit,
the region to the left and}
\centerline{below the blue curve is unstable.}

\pagebreak

\centerline{
\includegraphics{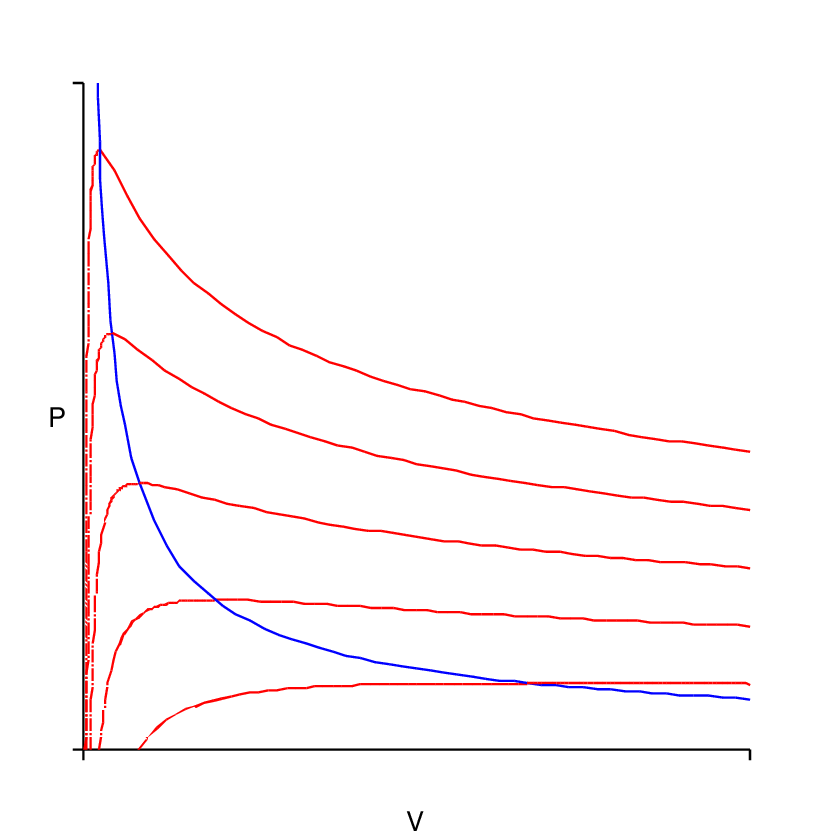}}
\vskip  10cm
\centerline{Figure 2: Black hole P-V diagram, showing curves of 
constant temperature}
\centerline{in AdS.  The blue line shows the stability limit,
the region to the left and}
\centerline{below the blue curve is unstable.}

\pagebreak

\centerline{
\includegraphics{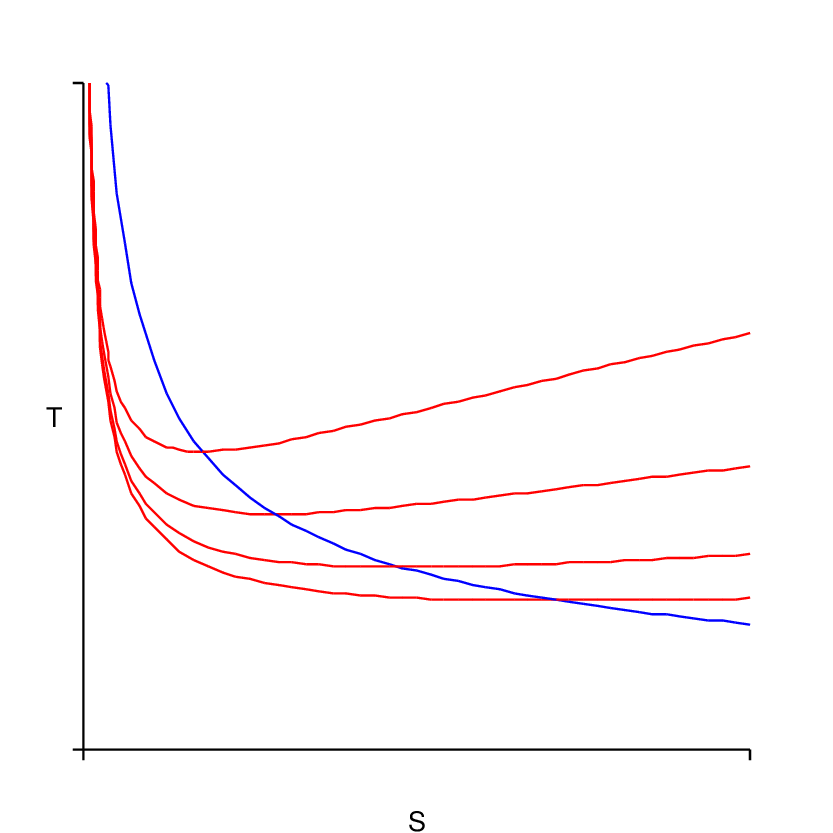}}
\vskip 8.8cm 
\centerline{Figure 3: T-S diagram, showing curves of constant pressure.}
\centerline{The region left of the blue line is unstable, temperature} 
\centerline{is
an increasing function of entropy in the stable region.}

\pagebreak

\centerline{
\includegraphics{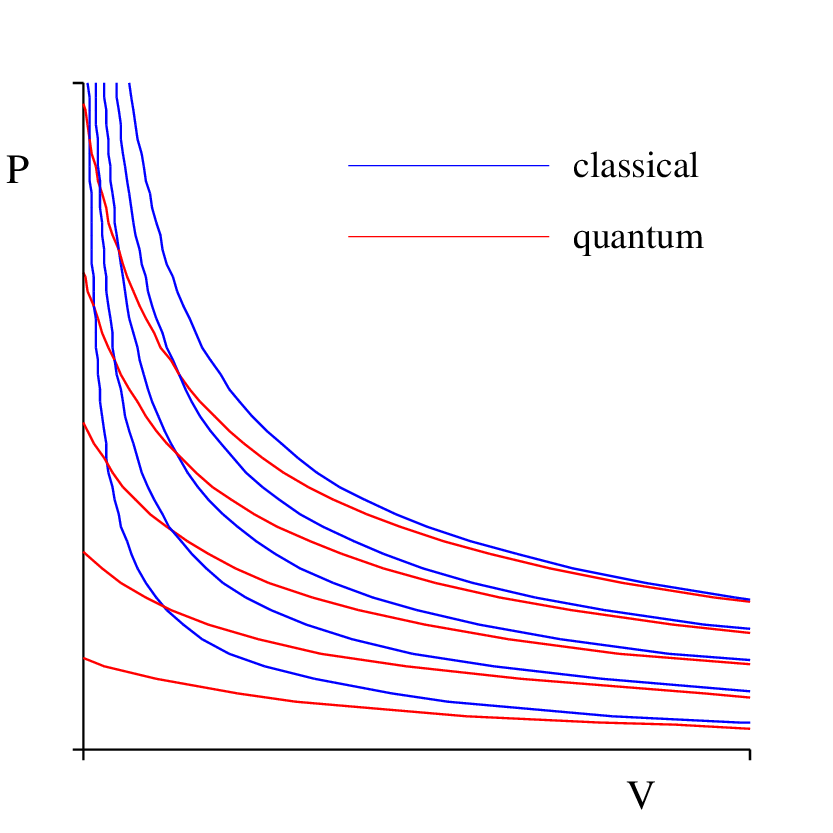}}
\vskip 10cm
\centerline{Figure 4: Black hole P-V diagram, showing curves of 
constant temperature}
\centerline{in AdS$_3$.  The blue lines show the classical curves, $P\propto \frac 1 {\sqrt{V}}$. Red lines show}
\centerline{the quantum corrected equation of state.
The region left of the blue line is unstable,} 
\centerline{temperature is
an increasing function of entropy in the stable region.}

\end{document}